\documentclass[pre,aps,twocolumn]{revtex4}
\usepackage{epsfig}

\def\be{\begin{equation}}
\def\ee{\end{equation}}
\def\ba{\begin{array}}
\def\ea{\end{array}}
\def\bea{\begin{eqnarray}}
\def\eea{\end{eqnarray}}
\def\d{{\mathrm d}}
\begin{document}

\def\Fig{Fig}
\def\FigSep{Fig}

\title{Universal depinning force fluctuations of an elastic line: \\
Application to finite temperature behavior}

\author{Damien Vandembroucq, Rune Skoe  and St\'ephane Roux}

\affiliation
{Unit\'e Mixte CNRS/Saint-Gobain ``Surface du Verre et Interfaces''\\
39 Quai Lucien Lefranc, 93303 Aubervilliers cedex, FRANCE}

\begin{abstract}

The depinning of an elastic line in a random medium is studied {\it
via} an extremal model. The latter gives access to the instantaneous
depinning force for each successive conformation of the line. Based on
conditional statistics the universal and non-universal parts of the
depinning force distribution can be obtained. In particular the
singular behavior close to a (macroscopic) critical threshold is
obtained as a function of the roughness exponent of the front. We show
moreover that the advance of the front is controlled by a very tenuous
set of subcritical sites.
Extension of the extremal model to a finite temperature is
proposed, the scaling properties of which can be discussed based on
the statistics of depinning force at zero temperature.
\end{abstract}
\pacs{PACS numbers }
\maketitle

\section{Introduction}

The depinning of elastic interfaces allows to describe the
phenomenology of a variety of physical phenomena such that the motion
of a wetting contact line on a disordered
substrate\cite{Moulinet-EJPE02}, the propagation of a fracture front
\cite{Schmittbuhl-PRL97} in a heterogeneous material or the advance of
a magnetic wall in a thin layer\cite{Lemerle-PRL98}. The depinning
transition can be seen as a non equilibrium critical phenomenon (see
Ref. \cite{Leschhorn-AnnPhys97,Kardar-PR98,Fisher-PR98} for recent
reviews). The system is driven by an equivalent external force (the
magnetic field for the domain wall, the stress intensity factor for
the fracture front...)  which plays the role of a control
parameter. The richness of the phenomenology arises from the
competition between the disordering effect of the quenched pinning
potential and the ordering (smoothing) effect of the elastic forces
acting on the front. The nature (short range/long range) of the latter
will directly affect the universality class of the depinning. Moreover
depending on the strength of the disorder, the motion front will
either be collective (weak pinning) or consists of successive
avalanches (strong pinning). In the following we focus on the latter
situation.

To study the pinning/depinning transition, a specific algorithm has
been introduced under the generic name of ``extremal model''
\cite{Sneppen-PRL92,Leschhorn-PRE94,Schmittbuhl-PRL95,Tanguy-PRE98}.
It consists in adapting the driving force so that only one site (the
weakest one) can depin at a time. In so doing, the system is compelled
to stay at the edge of the critical state (the front is driven at an
infinitely low velocity).  Thus the evolution consists in a series of
equilibrium positions, because the driving force is adjusted to meet
this condition.  It thus can be seen as an ideal quasi-static
driving\cite{Zapperi-BJP00}. However, the price to pay in such a
description, is that the dynamical aspects of the propagation are not
accounted for (see Ref. \cite{Narayan-PRE00} for details on this
subject). In a more complete picture either inertia or viscosity would
have to be introduced. As dynamics is scarifies, time metric is not
included in the ``extremal model'' description.  What is preserved is
simply the ordering of successive configurations.

Close but below threshold, at finite temperature, an Arrhenius
activation mechanism can allow for a steady ``subcritical''
propagation. Note that this situation differs from the creep regime at
very low forcing which has been extensively studied in recent years
\cite{Feigelman-PRL89,Marchetti-PRL96,Chauve-PRB00,Nattermann-AdvPhys00}. In
the latter, the driving force is very close to zero and thermal
activation may allow back and forth motion along the direction of
propagation. Close to threshold however, we may neglect the
probability of a backward motion.  Moreover in the context of
fracture or wetting, chemical reactions may produce an irreversible
motion of the front. In the following we will not consider the
possibility of front receding.  The activated mechanism thus turns out
to be relatively easy to implement as an extension of the extremal
model, where the latter is recovered in the limit of zero temperature.
The motion consists in a succession of fast moves interrupted by long
trapping events which require a thermal activation step to overcome an
energy barrier. If the temperature is very low compared to the trap
depths, the first site to depin will be the weakest one. In this limit
of a vanishing temperature the system will thus naturally follow an
{\em extremal} dynamics.  The degeneracy of the time metric can
however be clarified by studying the zero temperature limit.  This
description is valid for a subcritical forcing, close to threshold.
This forcing introduces a finite correlation length.  However, the
thermal activation itself introduces another finite correlation
length.  The competition between these two length scales will finally
control the scaling property of the propagation.

The aim of this paper is thus a careful study of the effect of
temperature on both the dynamics and the critical properties of
the depinning fronts.  A key parameter in the finite temperature
depinning is the ratio between the thermal energy and the trap
depths all over the front. The latter directly depend on the
fluctuations of the depinning forces at zero temperatures. The
external force needed to depin the front from a blocked
configuration is actually a highly fluctuating quantity.  The
knowledge of the distribution of these depinning forces thus appears to
be an essential ingredient in the study of depinning at finite
temperature.

The first part of this paper is thus devoted to the study of the
depinning forces at zero temperature.  We show in particular that
the distribution of depinning forces presents some universal
features close to the critical threshold. A criterion
characterizing the transition between low and high temperature
behavior requires additional information about the distribution of
trap depths along the front. Again we show that this distribution
presents some universal features. In the second part we introduce
a Monte-Carlo like algorithm allowing to drive the front under
constant forcing at finite temperature. We briefly present results
concerning the two limiting cases (at low and high temperatures)
and finally focus on the transition between the two regimes. The
role of the two competing correlation lengths governed
respectively by the driving force and the temperature will be
discussed in this last section.

\section{Depinning force distributions}

\subsection{Description of the model}

At each time step, the front is
represented as a single valued function $h(x)$, with the additional
assumption that its slope is small. The mean orientation of the front
is along the $x$ axis, while it propagates along a perpendicular
direction $y$. For numerical simulations, the front is discretized on
a regular grid of size $L$with periodic boundary conditions: $h_i=h(x_i)$.

Time is also discretized and incremented
by one unit at each elementary move of the front.  This time is thus a
simple way to order the successive events, but it does not correspond
to a physical time.  Additional information has to be introduced to
describe the off-equilibrium motion, and thus decide whether viscosity
or inertia, or activation processes control the dynamics.  The
strength and weakness of the extremal approach is that only successive
static positions are described, while the transition between these
conformations becomes part of the postulated rules of the model but do
not stand for a real dynamic.  Thus in this first part, $t$ should not
be misinterpreted as a real time.  However, in the finite temperature
section, we will come revisit this question and see how the extremal
dynamics can be recovered with a more realistic dynamics.

As introduced above,  the motion of the front driven at an
external force $F$ depends only on the competition between a local
trapping force $\gamma(x,h(x))$ and an elastic interaction
$F^{el}(x)$.  We now specify them in more details.

The distortion of the front due to the random
environment induces elastic restoring forces. In the case of wetting
the latter are the capillary forces.  Using a small slope
approximation, the elastic force contribution is linear w.r.t. $h(x)$,
$F^{el}(x)=\int G(x-x') h(x') \d x'$. Depending on the physical
situation considered, these interactions can be short or long
ranged. In the case of wetting the kernel $G$ presents an algebraic
decay $G(r)\propto r^{-2}$ up to the capillary length. In a Hele-Shaw
experiment however, the cut-off scale of the capillary forces is given
by the width of the cell and the Green function can be approximated by
$G\simeq \delta''$, the second derivative of the Dirac distribution
$\delta$. These short ranged interactions are then well described by a
Laplacian term. In the following we focus on the latter situation with
$F^{el}\propto -\partial^2 h/\partial x^2$, or in the discretized version,
$F^{el}_i=-h_{i-1}+2h_i-h_{i+1}$.

The randomness of the environment can be introduced in
two respects: first in the spatial distribution of the traps, second
in the distribution of the trap depths. No correlation is considered
here and these two quantities entirely characterized by their
statistical distribution. Note that the same critical behavior is
obtained as soon as either the trap positions or the trap depths are
random.  Either one of these two quantities can be a constant without
changing the universality class of the model, i.e. all critical
properties remain unaffected.  We have performed a number of different
numerical simulations changing the shape of the two distributions, and
the only changes which are observed concerns small scales in space and
time.  In the following we will consider that the trap depth
$\gamma(x,h(x,t))$ and the front advance at the active site are
uniformly sampled between 0 and 1.

In order to study the depinning force fluctuations, we drive the
system with an extremal dynamics. This consists in adapting the
external force so that only one site can depin at each iteration
step. For a given external force $F$, the criterion for depinning
at a particular position $i$ on the front is
\begin{equation}
F>F_i(t)=\gamma_i(h_i(t))+F^{el}_i(\{h_j(t)\})
\end{equation}

One can thus naturally define the extremal site $(i^*,h_{i*})$
such that $F_c(t)=\min_i F_i(t)=\gamma_{i*}+F^{el}_{i*}$ is the
minimal external force to apply so that at least one site of the
front can depin. The depinning criterion for a particular
conformation of the front is thus
\begin{equation}
F>F_c(t)=\min_i \left[ \gamma_i(h_i(t))+F^{el}_i(\{h_j(t)\}) \right]
\end{equation}

The front depinning force $F_c(t)$ is in our simple version
totally controlled by the front geometry.  The extremal dynamics
simply consists in tuning at each iteration step $t$ the external
force at exactly the depinning force of the current front
conformation : $F(t)=F_c(t)$. From the dynamical point of view it
corresponds to drive the front at vanishing velocity with a very
stiff spring. The system thus remains constantly at the edge of
the critical state. The front explores a large number of different
configurations to which correspond fluctuations of the depinning
force $F_c(t)$. By definition the critical threshold $F^*$ is the
minimum constant force to apply so that the front can advance
indefinitely, it thus corresponds to the maximum of the front
depinning force $F_c$ over all conformations:
 \begin{equation}
 F^*=\max_t F_c(t)=\max_t\left[\min_i
 \left[\gamma_i(h_i(t))+F^{el}_i(\{h_j(t)\})\right]\right]
 \end{equation}

From the numerical point of view, the extremal dynamics finally
amounts to iterate the following sequence:{\it i)} identify the
extremal site $i^*$; {\it ii)} advance the front at position $i^*$ by
a random increment; {\it iii)} 
update the trap depth, $\gamma_{i*}$ on the site $i^*$ corresponding
corresponding to the new front position $h(i^*)$; {\it iv)} update the
elastic couplings $F^{el}_i$ on the front to account for this local
advance. The most time consuming step is the identification of the
extremal site.  Numerically efficient implementation can be used to
reduce the computation cost of each iteration to $\log_2(L)$
operations because of the short range nature of the interaction (note
that in case of long range interactions, introducing an ultrametric
distance along the front \cite{VR-preprint03} allows to reach the same
numerical efficiency without changing the universality class of the
model).


\begin{figure}[t]
\epsfig{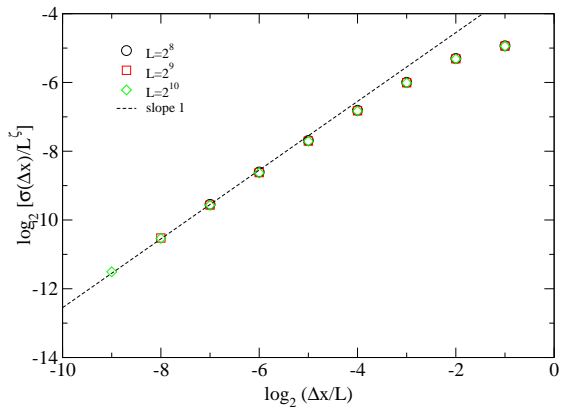}
\caption{\label{ecz2} Standard deviation $\sigma(\Delta x)$ of the
height difference between points separated by a distance $\Delta x$ for
different system sizes $L=2^8,2^9,2^{10}$.  The rescaling $x\to x/L$,
$\sigma \to \sigma/L^{\zeta-1}$ with $\zeta=1.25$ allows to capture
the finite size effects.}
\end{figure}

\subsection{Scaling properties}

Let us summarize the expected scaling behaviors for an elastic line
driven at a constant force $F$ close to (but below) threshold $F^*$.

\begin{itemize}
\item Below threshold $F^*$ the length of the typical advance
$\xi_\perp$ scales as $\xi_\perp \propto (F^*-F)^{-\nu_\perp}$

\item The correlation length along the front scales as
$\xi_\parallel\propto |F^*-F|^{-\nu_\parallel}$.

\item The relaxation time $\tau$ of a line segment of length $\xi$
scales as $\tau \propto \xi_\parallel^z$; $z$ is called the dynamic
exponent.

\item At threshold, in the steady state, the front conformation is
self-affine, characterized by a roughness exponent $\zeta$.  This can
be shown by studying the average power spectrum of the front which
scales as $k^{-1-2\zeta}$. Similarly the wavelet transform also
reveals the same exponent\cite{Simonsen-PRE98}. In real space, the
typical width $w$ (r.m.s. height) of the front over an interval
$\Delta x$ scales as $w (\Delta x)\propto (\Delta x)^\zeta$, when
$\zeta$ is smaller or equal to 1. This is observed for some long-range
elastic kernel. In our case, where the elastic kernel is the local
curvature, we will see below that the $\zeta$ exponent is larger than
unity. In this ``super-roughening'' case, the scaling is
anomalous\cite{Lopez-PRE97} and the previous relation has to be
corrected to $w (\Delta x)\propto L^{\zeta-1}\Delta x$. On
Fig. \ref{ecz2} we show numerical evidence for the validity of this
``super-rough'' scaling with the value $\zeta=1.25$ for the roughness
exponent.

\item Close to the critical threshold the motion consists of
successive localized avalanches of lateral size $\Delta x$. The size
distribution of these jumps follows a power law up to the correlation
length $\xi_\parallel$. Below $\xi_\parallel$, the probability of
observing jumps of size $\Delta x>\ell$ scales as $P(\Delta x>\ell)
\propto \ell^{-1-A}$.  In the framework of an extremal dynamics
driving, the driving force is not constant but is adjusted at the
current depinning force $F_c(t)$. An avalanche triggered at a force
$F$ thus corresponds to a sequence of depinning events such that the
current depinning force remains below the driving force $F_c(t)<F$.
The distributions of avalanches can thus be directly derived from the
fluctuations of this depinning force $F_c(t)$. A directly related
quantity is the probability $P(\Delta x|\tau')$ that after a sequence
of $\tau'$ events the depinning site has moved  a distance $\Delta
x$. We have
\begin{equation}
P(\Delta x|\tau') \propto \Delta x^{-a} \phi\left(\frac{\Delta
x}{\tau'^{1/z'}}\right)
\end{equation}
where $\phi(u) \propto u^a$ for $u \ll 1$ and $\phi(u) \approx
\mathrm{cste}$ for $u \gg 1$. In the framework of extremal dynamics,
$z'$ corresponds to a dynamic exponent (see
Ref. \cite{Krishnamurthy-EPJB00,Narayan-PRE00} for details on the
relations between $z'$ and the genuine dynamic exponent $z$
introduced above. In the following we restrict the study to extremal
dynamics and we use the notation $z$ for the sake of simplicity.
\end{itemize}

\begin{figure}[t]
\epsfig{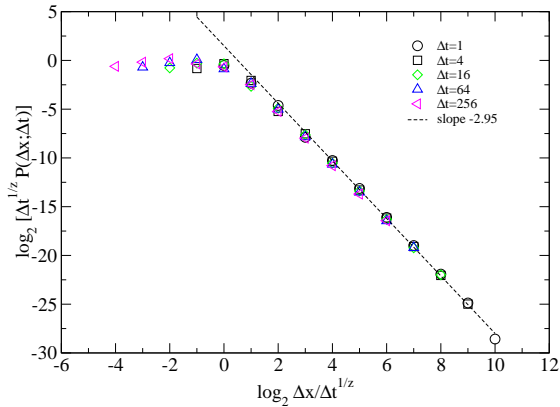}
\caption{\label{furuberg} Distribution of distances between active
sites corresponding to depinning events separated by a time delay
between $\Delta t=1$ and $\Delta t2^5$. After rescaling all
distributions fall on a master curve.  This rescaling has been
obtained with the value $z=2$ for the dynamic exponent. The dashed
line corresponds to a power law of exponent $a=-2.95$}
\end{figure}

Simple scaling relationships immediately derive from these
properties and allow to reduce the number of independent exponents.

\begin{itemize}

\item Below threshold the front advance is confined to a region of
length $\xi_\parallel$, and in the case of a roughness exponent $\zeta
>1$ the mean advance over this region is of order $\xi_\perp \propto
\xi_\parallel L^{\zeta-1} \propto
(F-F^*)^{-\nu_\parallel}L^{\zeta-1}$. This leads to
$$
\nu_\perp=\nu_\parallel=\nu
$$ with an unusual size-dependent prefactor. Note that this property
is only true for a roughness exponent $\zeta>1$, otherwise
$\nu_\perp=\nu_\parallel \zeta$.

\item Over a region of extent $\xi_\parallel$, the typical macroscopic
curvature is of order $\xi_\parallel^{\zeta-2}$, and the latter
scaling should balance the fluctuation of depinning force observed in
order to depin such a segment: $\Delta F\propto
\xi_\parallel^{-1/\nu}$.  Therefore\cite{Kardar-PR98}
\begin{equation}
\label{elastic-scaling}
\zeta +\frac{1}{\nu} = 2
\end{equation}

\item Let us recall that time only counts individual motions, and is not
physical.  Therefore the dynamic exponent is an improper
denomination.  Nevertheless, in this section, we keep this
denomination since it refers to its common usage in statistical
models.

The mean advance of a segment of length $\xi$ scales as
$L^{\zeta-1}\xi$ and the size of an avalanche of lateral extension
$\xi$ writes $\tau \propto \xi^2 L^{\zeta-1}$ which leads
immediately to
\begin{equation}
z=2
\end{equation}
whenever $\zeta>1$. Note however that even if the dynamic exponent
$z=2$ appears to be super-universal as soon as $\zeta>1$, the relation
between $\tau$ and $\xi$ becomes system size dependent.  Otherwise,
for $\zeta<1$, the same argument leads to
$z=1+\zeta$\cite{Tanguy-PRE98,Krishnamurthy-EPJB00}.

\end{itemize}

All exponents but $a$ which characterizes the avalanche behavior can
then be directly derived from the roughness exponent $\zeta$. An
simple scaling relationship can however be established in the context
of the avalanches\cite{Krishnamurthy-EPL00}.  Let us start from the
probability $P(\Delta x|\tau)$ that after a sequence of $\tau$ events
the depinning site has moved from a distance $\Delta x$. The distance
$\Delta x$ is nothing but the sum of all successive jumps $\Delta y_i$
occurring at steps $i \in [1,\tau]$. The distribution of these jumps
being $P(\Delta y) \propto (\Delta y)^{-a}$. Assuming no time
correlation we can apply a generalized central limit theorem for the
sum variable $\Delta x$. The value of the exponent $a$ being slightly
below $3$, the limit distribution is not Gaussian but a L\'evy
distribution ${\cal L}_{a-1}$ that exhibits at infinity the same power
law behavior as the original power law distribution:${\cal
L}_{a-1}(\Delta x)\propto (\Delta x)^{-a}$:

\be P(\Delta x,\tau) \approx \frac{1}{\tau^{\frac{1}{a-1}}} {\cal
L}_{a-1} \left( \frac{\Delta x}{\tau^{\frac{1}{a-1}}} \right) \ee
which leads to the scaling relationship $a=1+z$. In the present case
of Laplacian elasticity, $z=2$ and the prediction $a=3$ slightly
overestimates the value obtained by numerical simulations $a\approx
2.95$ (see Fig. \ref{furuberg}. This good agreement can be interpreted
as the quasi absence of temporal correlations. Note however that in
the case of long range elasticity where the role of temporal
correlations is expected to be higher, this scaling relationship
becomes $a=2+\zeta$ which is far above the measured values $a\approx
\alpha$ where $\alpha$ is the exponent of the elastic redistribution
function. In the latter case, temporal correlations become more
significant.


\begin{figure}
\epsfig{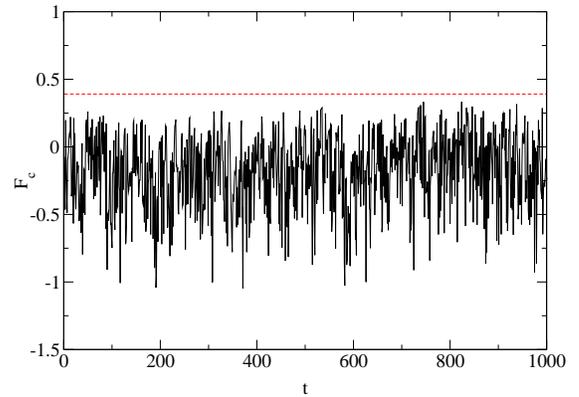}
\caption{ \label{Fc_snapshot} Sequence of depinning forces
$F_c(t)$. The dashed line corresponds to the value of the critical
threshold $F^*=\max_t F_c(t)$}
\end{figure}

\subsection{Scaling and universality of depinning forces fluctuations}

\begin{figure}[t]
\epsfig{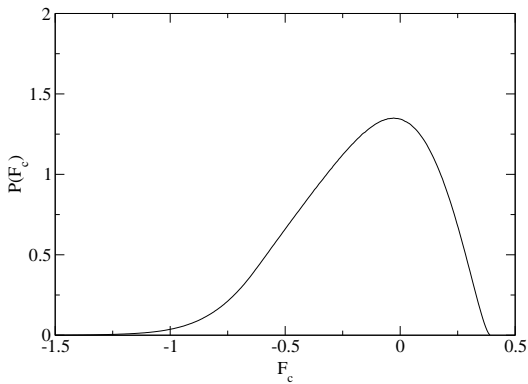}
\caption{\label{Fc_distribution_raw} Distribution of depinning
forces, $p(F_c)$. }
\end{figure}

On Fig. \ref{Fc_snapshot} we display a sequence of depinning forces
$F_c(t)$ observed over $1000$ steps of an extremal dynamics simulation
and the corresponding distribution $P(F_c)$ in
Fig. \ref{Fc_distribution_raw}.  The upper force value of this
distribution corresponds to the critical threshold $F^*$. In the
following we will argue that close to the critical threshold the
distribution of depinning forces exhibits a universal behavior:
\begin{equation}\label{eq:defmu}
P(F_c) \propto (F^*-F_c)^\mu
\end{equation}


\begin{figure}[t]
\epsfig{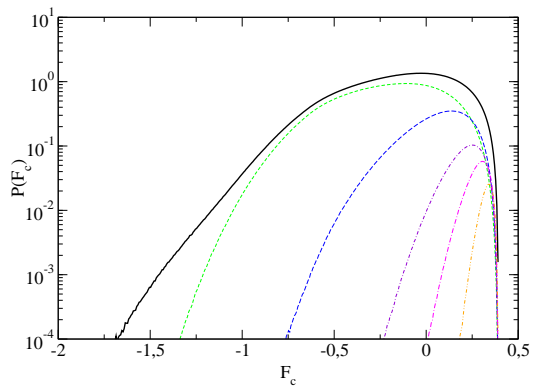}
\caption{\label{Fc_distribution_d} Distribution of depinning forces
conditioned by their distance to threshold, $p(F_c,d)$, and the
resulting global distribution $P(F_c)$ (bold). The larger the distance $d$
between consecutive active sites, the narrower the distributions and
the closer to the critical force.}
\end{figure}

Let us come back now to the series of depinning events.  The
depinning force $F_c$ fluctuates and can be described by a
statistical distribution. The maximum value of the latter gives
the value of the critical threshold. In the framework of a
simulation this threshold $F^*$ is obtained as the maximum
depinning force over the ensemble of depinning events. Restricting
ourselves to a finite system of lateral extension $L$, our
determination will be biased by finite size effects. This invites
us to introduce a simple prescription to get a very accurate
determination of $F^*$.  Conditioning the distribution of
depinning force by the distance $d$ between consecutive active
sites, we can built distributions $P(F_c;d)$. We expect these
distributions to converge to a Dirac distribution centered at $F^*$
as $d$ diverges since we probe only critically pinned configurations. This effect is directly observed on Fig.
\ref{Fc_distribution_d} where we superimpose the original
distribution of depinning forces $P(F_c)$ and the contributions
corresponding to increasing jump sizes $d$. We observe that,
except for small jumps the successive distributions keep a similar
shape up to scaling factors. In the following we check that these
conditional distributions can indeed be rescaled onto a unique
master curve using the scaling:
\begin{equation}
P(F_c;d) = \frac{1}{d^{-b}}\psi\left( \frac{F^*-F_c}{d^{-b}} \right)
\end{equation}

\begin{figure}[t]
\vspace{0.6cm}
\epsfig{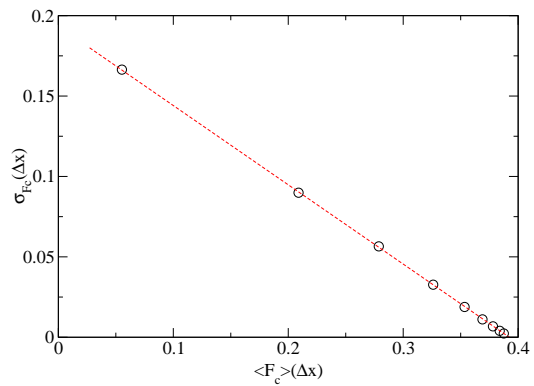}
\caption{\label{meanFc_vs_sigmaFc} The linear relationship between
the mean depinning force and its fluctuation allows to extrapolate
the value of the depinning force for an infinite distance between
consecutive active sites {\it i.e.} the depinning threshold $F^*$
for an infinite system.  A numerical fit gives $F^*=0.392\pm
0.002$.}
\end{figure}

As a direct consequence of such a scaling we should in particular
observe that the first and second cumulant of these distributions
{\it i.e} the difference $F^*-\langle F_c\rangle$ and the standard
deviation both scale as $d^{-b}$:
\begin{eqnarray}
\delta F_c(d)&=&F^* - \langle F_c \rangle (d) \propto d^{-b} \;,\\
\sigma_{F_c}(d)&=&\sqrt{\langle F_c^2 \rangle (d)-\langle F_c \rangle^2 (d)}
\propto d^{-b}
\end{eqnarray}
This scaling directly leads to a linear dependence between
$\langle F_c(d)\rangle$ and $\sigma_{F_c}(d)$. On Fig.
\ref{meanFc_vs_sigmaFc}, we check indeed that for large values of
the jump size $d$, this linear behavior is obeyed. By
extrapolation of this linear relationship for an infinite system
for which the width $\sigma_{F_c}(d)$ cancels, we obtain a precise
determination of the critical threshold $F^*$. We obtain in the
present case 
\begin{equation}
F^*=0.392\pm0.001
\end{equation}


\begin{figure}[b]
\vspace{0.5cm}
\epsfig{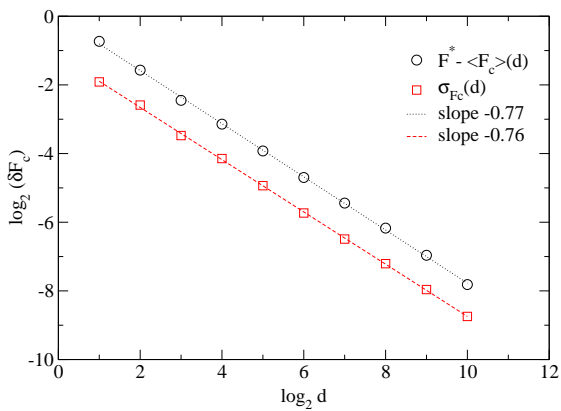}
\caption{\label{Fc_scaling}The mean distance from the depinning force
to the depinning threshold and the depinning force fluctuations
calculated for given distances between consecutive active sites obey a
scaling law with the latter distance. The straight lines plotted as a
guide to the eye correspond to power laws of exponent $0.76$ and
$0.77$ consistent with the expected value $2-\zeta \approx 0.75 $.}
\end{figure}

Using the latter value we plot on Fig. \ref{Fc_scaling} the evolution
of $\delta F_c(d)$ and $\sigma_{F_c}(d)$ versus $d$ in logarithmic
scale. The observed linear behavior corresponds to power laws of
exponent $b=0.75$. The value of the latter exponent can be understood
since the depinning force $F_c$ directly derives from the local
curvature.  The latter can be estimated using the elastic kernel and
the roughness of the front. In the present case we have $\delta
F^{el}(d) \propto d^{\zeta-2}$ where $\zeta \approx 1.25$ is the
roughness exponent of the depinning front.  This estimation
$b=2-\zeta\approx 0.75$ is consistent with our numerical
data. Moreover we can identify $1/b$ to the exponent $\nu$
characterizing the divergence of the correlation length close to
threshold and we recover the scaling law of Eq.
(\ref{elastic-scaling}):
\begin{equation}
\zeta+\frac{1}{\nu}=2
\end{equation}


\begin{figure}[t]
\epsfig{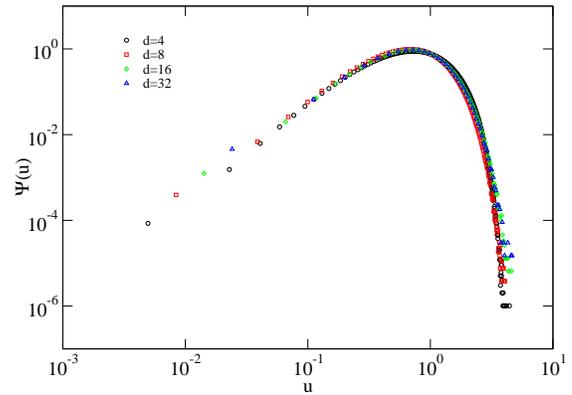}
\caption{\label{universal} After rescaling the distributions of
depinning forces corresponding to distances $d=4,8,16, 32$
collapse onto a unique universal master curve. 
} 
\end{figure}

Using this scaling we check on Fig. \ref{universal} that the
successive conditional distribution collapse onto a single
universal master curve. The knowledge of the distribution of
distances between successive depinning sites allows us now to
derive the behavior of the force distribution close to 
threshold.
\begin{eqnarray}
P(F^*-F_c) &= &\int  P(F_c;\ell) p(\ell) \d\ell \\
\nonumber
&=&\int \ell^{\frac{1}{\nu}} \psi\left[(F^*-F_c)
\ell^{\frac{1}{\nu}}\right] \ell^{-a} \d\ell\\
\nonumber
&=&\left(F^*-F_c\right)^{-1+a\nu-\nu} \int u^{1-a\nu +\nu-1} \psi(u) \d u\\
\nonumber
&\propto&\left(F^*-F_c\right)^{(a-1)\nu-1}
\end{eqnarray}
We thus obtain a universal behavior of the distribution of
depinning forces $F_c$ close to  threshold as
presented in Eq.~\ref{eq:defmu}, with
 \be
 \mu=(a-1)\nu-1\approx 1.6
 \ee

Note in addition that when the distance $d$ is small compared to the
correlation length $\xi$, $x\equiv d/\xi\ll 1$, the dependence of the
$P$ distribution w.r.t. $(F^*-F_c)$ should be independent of $d$, so
that $P(F_c;d)$ and $P(F_c)$ share the same behavior (see
Fig.\ref{Fc_distribution_d}).  Hence, we expect $\psi(x)\sim x^\mu$
for $x\ll 1$, in good agreement with the behavior shown on
Figure~\ref{universal}.


At this point it is important to clarify that the distribution $P$ by
itself is not universal and depends on the local definition of the
front advance and the trap depth. The details of the random
distributions are absorbed in the small time and scale behavior.  In
contrast collective effects contribute to large jumps in the active
sites and $\Psi$ is universal as well as the singular behavior of $P$
close to threshold.

\subsection{Subcritical sites along the depinning front}

\begin{figure}[tbp]
\epsfig{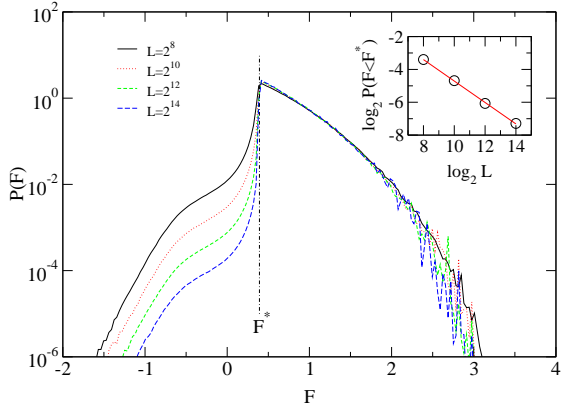}
\caption{\label{histo_f} Distribution of individual depinning forces
$F_i=\gamma_i-F^{el}_i$ along the front. The subcritical part of the
distribution $(F<F^*)$ is size dependent. In the insert we see that
the relative weight of this subcritical part follow a power law of
exponent $0.65$ with the size $L$. This suggests that the set of
subcritical site is fractal with a dimension $D=1-0.65=0.35$}
\end{figure}

In Fig. \ref{histo_f} we show the distribution of all individual
depinning forces $F_i=\gamma_i-F^{el}_i$ along the front obtained from
extremal dynamics. We see clearly that two different populations of
sites can be distinguished. Above threshold $F^*$, ${\cal Q}(F>F^*)$
follows some kind of truncated Gaussian distribution.  We also note
that there is a small but significant number of sites which carry a
force smaller than threshold, $F<F^*$.  Moreover, the decay of
frequency is very abrupt right below $F^*$.  We will argue that this
decay can be rationalized as a power-law of the difference
$(F^*-F_c)$, ${\cal Q}(F<F^*)\propto A(L)(F^*-F)^{-\varepsilon}$. In
the insert of Fig. \ref{histo_f} we plotted the proportion of
subcritical sites ${\cal Q}(F<F^*)$ against the system size. This
evolution appears to follow a power law ${\cal Q}(F<F^*)\propto
L^{D-1}$ with $D\approx 0.35$. We show below that $D$ can be regarded
as a fractal dimension of the set of subcritical sites along the
front.

\begin{figure}[b]
\centerline{\epsfxsize=.85\hsize \epsffile{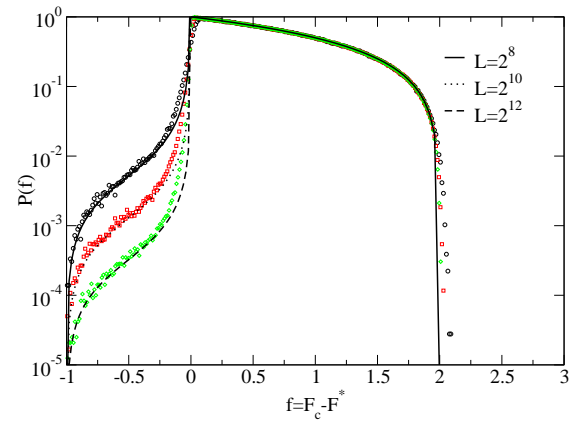}}
\caption{\label{fi:mf} Comparison between the result of a direct
numerical simulation (symbols) and the previous analytic expressions
(continuous curves).  The system size was set to $L=2^8,2^{10}$ and
$2^{12}$. As above we see that only the subcritical part is size
dependent}
\end{figure}

 In the Appendix, we
will introduce a mean-field version of the model where such a
distribution can be computed analytically.  In this simple variant of
the depinning model, the elastic force is not transfered from the
active site to its nearest neighbors but to two sites chosen randomly
along the front.  Although a quantitative comparison can not be
established because of the mean-field nature of the solvable model, on
a qualitative ground, the main features of the force distribution are
recovered. On Fig \ref{fi:mf}, one can distinguish again between a
Gaussian-like distribution for depinning forces above the critical
threshold and a singular part below the threshold. Moreover it can be
shown that the amplitude of this singular part is system size
dependent.

The ``subcritical'' sites which carry a force smaller than threshold,
 play a very important role, as we now try to clarify. At each time
step, the extremal site corresponds to the minimum force and always
remains smaller than $F^*$.  Thus the active site is always part of
the subcritical sites. Its depinning induces a local unloading and an
additional re-loading on its neighbors. We note that a subcritical
site has a chance of being activated wherever it is located, while a
site carrying a force larger than threshold can only become active if
one of its neighbors become active itself. These sites can thus be
regarded as a population of ``precursors'' of depinning.

In order to determine the distribution of the depinning forces
along the front we use the same tools as above. We  first show that
the distribution of $F_i$ along these sites depends on their
distance to the extremal one as a power law: ${\cal Q}(F_i<F^*) \propto
|x_i-x_{i^*}|^{-a'}$. We observe then that the conditional
distribution ${\cal Q}(F_i<F^*;|x_i-x_{i^*}|=d)$ obeys the same scaling
as the one studied above for the distribution of depinning forces.
This allows finally to estimate the power law behavior of the
distribution of the depinning forces of this population of
subcritical sites close to  threshold.

\begin{figure}
\epsfig{file=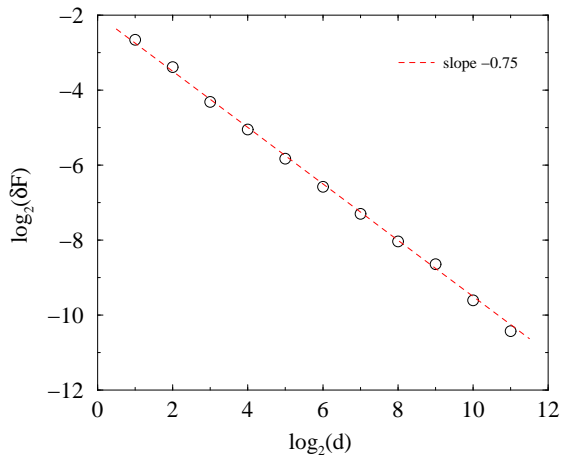, width=0.85\hsize}
\caption{\label{sumfdx} The difference $\delta F$ (current force
to threshold) of precursory sites located at a distance $d$ from
the extremal site scales as a power law of $d$. The straight line
corresponds to a power law of exponent $0.75\approx 2-\zeta$.}
\end{figure}

\begin{figure}[tbp]
\centerline{\epsfxsize=.85\hsize \epsffile{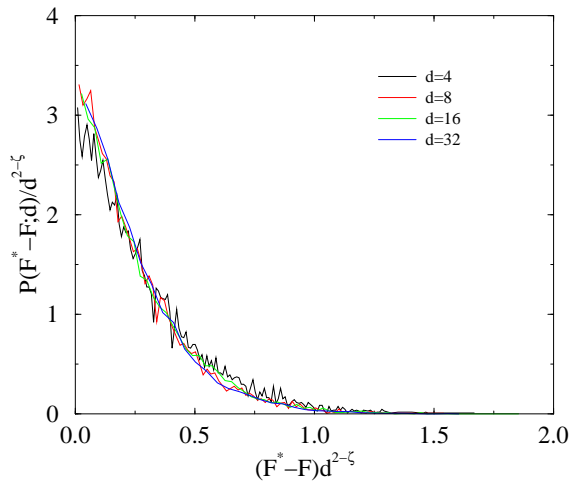}}
 \caption{\label{universal_F} After rescaling
the distributions of subcritical depinning forces $F_i$ conditioned
by the distance $d$ to the extremal site $i^*$, collapse onto a
single master curve.}
\end{figure}

On Fig. \ref{sumfdx} we plot in logarithmic scale the mean
difference to threshold $\langle F^*-F_i; |x_i-x_{i^*}|=d)\rangle$
for sites at a distance $d$ from the extremal one $i^{*}$. We
obtain the same scaling as for the distribution of the depinning
force $F_c$ conditioned by the distance between successive
depinning sites:
\begin{equation}
\langle F^*-F_i; |x_i-x_{i^*}|=d)\rangle \propto
d^{-\frac{1}{\nu}} \;,\quad \frac{1}{\nu}=2-\zeta
\end{equation}
and again we see on Fig. \ref{universal_F} that we can rescale all
conditioned distributions onto a single master curve provided the
distance $d$ is large enough.
\begin{equation}
P\left( F^*- F_i=f>0; |x_i-x_{i^*}|=d\right) =
d^{\frac{1}{\nu}}\varphi \left(fd^{\frac{1}{\nu}}\right)
\end{equation}

\begin{figure}[b]
\centerline{\epsfxsize=.8\hsize \epsffile{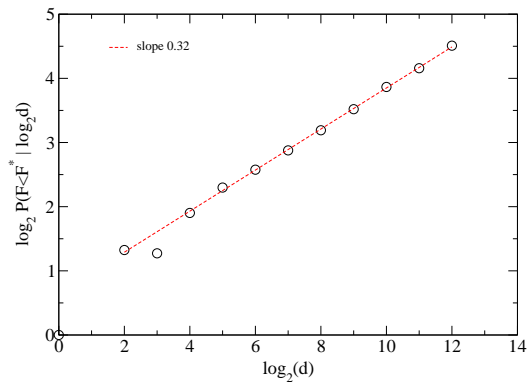}}
\caption{\label{nb_fdx} Distribution $n(d)$ of distances $d$ of
subcritical sites to the extremal site. The straight line
indicates a power law of exponent 0.32. The data being obtained
with a logarithmic sampling of $d$, this corresponds to a power
law $n(d)\propto d^{-c}$, $c\approx 1-0.32\approx 0.68$}
\end{figure}

On Fig. \ref{nb_fdx} we show that the subcritical sites are not
homogeneously distributed along the front. We plot in logarithmic
scale the distribution of distances $d$ along the front between
the extremal site and any other subcritical site. The straight
line obtained in logarithmic scale indicates that the latter obeys
a power law:
\begin{equation}
n\left(|x_{i^*}-x_i|=d ; F<F^* \right) \propto d^{-c}\;,\quad
\end{equation}
with $c\approx 0.68$, what indicates that the subcritical sites form a
fractal set of dimension $D=1-c\approx 0.32$, which is consistent with
the value $D\approx0.35$ obtained above for the evolution of the
population of subcritical sites against the system size.

\begin{figure}[tbp]
\epsfig{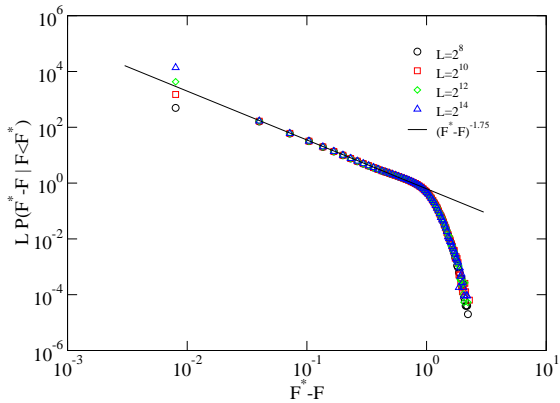}
\caption{\label{collapse_f} Distribution of subcritical depinning
forces $F<F^*$ along the front. Close to threshold the distribution of
subcritical sites follow a power law of exponent $-1.75$. }
\end{figure}

The above scaling relationships can now be used to estimate the
distribution of depinning forces along the front close to threshold:

\begin{eqnarray}
\label{P(Fi)}
&&Q(F^*-F_i=f>0) \\
\nonumber
&= &A(L)\int Q(F^*-F_i;\ell) n(\ell) \d \ell \\
\nonumber
&=&A(L)\int
\ell^{\frac{1}{\nu}} \phi\left[(F^*-F_i)\ell^{\frac{1}{\nu}}\right]
\ell^{-c} \d \ell\\
\nonumber
&=&A(L)\left(F^*-F_i\right)^{-1+c\nu-\nu} \int u^{1-c\nu+\nu-1} \psi(u) \d u\\
\nonumber
&\propto&A(L)\left(F^*-F_i\right)^{-\varepsilon}
\end{eqnarray}
with $\varepsilon=1+\nu D\approx 1.45$.  The size dependent prefactor
can be estimated using the fact that for a front of size $L$ the
typical cut-off of the elastic force is $L^{-1/\nu}$. Rewriting the
total number of sub-critical sites along the front then gives:
\begin{eqnarray}
{\cal N}(L)&=&
L A(L)\int^{L^{-\frac{1}{\nu}}}_{-\infty} \left(F^*-F_i\right)^{-\varepsilon}\\
&=&L A(L) L^{-\frac{1}{\nu}(1-\varepsilon)}\\
&=&A(L) L^{1+D} 
\end{eqnarray}
The fractal dimension of this population of subcritical sites being
$D$, we have in addition ${\cal N}(L)\propto L^D$ which imposes
 $A(L)\propto L^{-1}$ and we have finally (see Fig. \ref{collapse_f}):

\be
Q(F^*-F_i=f>0) \propto \frac{1}{L} \left(F^*-F_i\right)^{-\varepsilon}\;,\quad
\varepsilon=1+\frac{D}{2-\zeta}
\ee

We see thus that the depinning is controlled by the evolution of a
very tiny fraction of subcritical sites. These sites are distributed
fractally along the front and can be regarded as precursors of the
avalanches to come. Close to threshold the depinning force associated
to these sites is power law distributed with an exponent depending
both on the roughness exponent of the front $\zeta$ and the fractal
dimension $D$ characterizing the population of subcritical sites.

\section{Finite temperature dynamics}


We now consider the case of depinning at finite temperature. The
transition loses its critical character. As a matter of fact, even at
arbitrary low external driving, the front keeps moving. The velocity
is however extremely low, most of the time being spent to thermally
depin from blocked conformations. This creep motion has been
intensively studied in recent years from the experimental
\cite{Lemerle-PRL98} and theoretical points of view
\cite{Feigelman-PRL89,Marchetti-PRL96,Chauve-PRB00,Nattermann-AdvPhys00}. It
has been argued that at very low forcing, the velocity follows a
stretched exponential behavior against the driving force:
 \be v(F,T)
\propto \exp \left[ -\left( \frac{F^*}{F} \right)^m \frac{T}{T_0}
\right] 
\label{creep}
\ee
where $m=(2\zeta-1)/(2-\zeta)$. To our knowledge, the anomalous force
 scaling predicted in Eq. (\ref{creep}) has not been reproduced so far
 by numerical simulations\cite{Marchetti-PRB95,Roters-PRE01}.

Note again that such a
behavior is only expected to happen at very low forcing. This means in
particular that the macroscopic motion of the front results from a
balance between microscopic backward and forward motions.

In this section we focus our study on a slightly different situation,
namely, a driving force in the vicinity of the critical
threshold. This allows to neglect the probability of a backward
motion.  At finite temperature the motion consists of a series of
rapid finite moves interrupted by long pinned stages from which the
front can only escape by thermal activation.  Two limit cases can
actually be distinguished regarding the thermal depinning of a blocked
conformation of the front. At very low temperature, the front most
likely depins at the extremal site. Thus the sequence of depinning
events is not affected by the introduction of a temperature. The
computation of the time spent by the front to escape any blocked
conformation allows however to reintroduce a physical time in this
quasi extremal dynamics. The knowledge of the depinning force
distributions $P(F_c)$ thus serves to quantify the velocity of the
front. The second limit case corresponds to high temperatures. During
a transient, all sites are free to advance; then the curvature of the
front becomes comparable to the temperature and the dynamic of the
front is rather well described by an annealed dynamics. This induces a
scaling of the front roughness different from the one obtained at zero
temperature: we observe a roughness exponent $\zeta_T\approx 0.5$
instead of the zero temperature value $\zeta \approx 1.25$.

In the following we first present the simple Monte Carlo algorithm
used to run the dynamics of the front at finite temperature. The
latter is obtained by allowing activated depinning along the front. We
then give results obtained in the low and high temperature limits.  We
finally present a criterion allowing to discriminate between these two
regimes. It appears actually that beside the ``static'' correlation
length $\xi$ of the critical transition another characteristic length
of thermal origin $\xi_T$ plays a key role in the problem: below
$\xi_T$ the front follows a $T=0$ dynamics but above $\xi_T$ the
activated depinning become dominant and the front follows an annealed
Edwards-Wilkinson like dynamics\cite{Barabasi-book95}.


\begin{figure}[tbp]
\epsfig{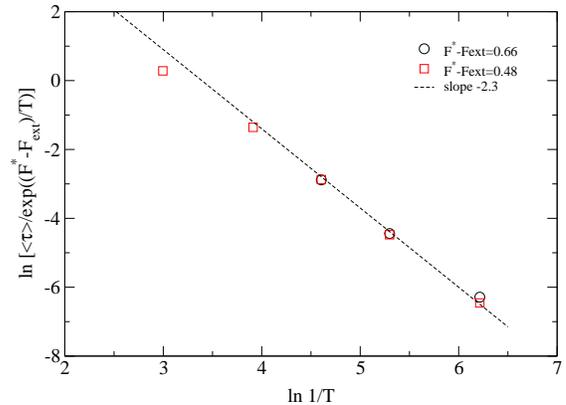}
\caption{\label{act-ext} Dependence of the mean waiting time on the
temperature after rescaling by the dominant Arrhenius factor. The
straight line indicates a power law behavior of exponent $2.3$ to
compare with the value $(a-1)\nu\approx 2.6$ obtained in
Eq. \ref{ext-act} for the low temperature dynamics.}
\end{figure}

\subsection{A simple Monte-Carlo algorithm for depinning at finite T}

Let us come back to the definition of the dynamics. Consider a
point $i$ along the front. At zero temperature the criterion for
depinning at this particular site is:
\begin{equation}
\label{prop_crit} D_i=F^{ext} + F^{el}_i  -\gamma_i> 0
\end{equation}

Note here that for the sake of simplicity we only consider the
fluctuating part of the random traps, the constant part, call it
$\Gamma$ being implicitly absorbed in the external driving force:
$F^{ext}=F^{ext}_{(0)}-\Gamma$. The values of this effective driving
force may thus be arbitrarily low and even negative without violation
of the hypothesis of microscopic forward motion mentioned above.

At finite temperature we can estimate the local depinning time
$t_i$ {\it via} an Arrhenius term:
\begin{equation}
\label{Arrhenius}
p(t_i)=\frac{1}{\tau_i}\exp\left(-\frac{t_i}{\tau_i}\right)\;, \quad
\tau_i= \frac{1}{\nu_0}\max\left[1;\exp\left(\frac{-D_i}{T}\right)\right]
\end{equation}
where $\nu_0$ corresponds to a natural frequency of vibration at
the microscopic scale.

From the numerical point of view we can use the following
algorithm. At time $t$, all points satisfying the criterion of
propagation (\ref{prop_crit}) are advanced simultaneously up to the
next traps. The waiting times $\tau_j$ corresponding to the points $j$
that do not satisfy the criterion are computed from their respective
distributions. Points such that the waiting time is below the time
unit are also advanced. The elastic forces are then updated and time
is incremented. If no point along the front has been 
advanced, the location of the first site to thermally depin and the
associated waiting time are drawn from the Arrhenius
distributions. Again the selected point moves forward, the elastic
forces are updated along the front but the time is incremented by the
waiting time spent in this blocked conformation. Let us now  be more
explicit about the determination of the waiting time and the location
of the depinning site in the case of a blocked conformation.

The probability $P_i(t)dt$ that the site $i$ is the first to
thermally depin in the time interval $[t,t+dt]$ can be written:
\begin{eqnarray}
P_i(t)&= &\frac{1}{\tau_i}\exp\left(-\frac{t}{\tau_i}\right)
 \prod_{j\ne i} \int_t^{\infty}
\frac{1}{\tau_j}\exp\left(-\frac{u}{\tau_j}\right) \d u\\
&=& \frac{1}{\tau_i} \exp\left(-\frac{t}{\tau^*}\right)\;,\quad
\frac{1}{\tau^*}=\sum_j \frac{1}{\tau_j}
\end{eqnarray}

The probability $r_i$ that the activated depinning event takes
place at this particular site $i$ is thus independent of time
\begin{equation}
r_i=\int_0^\infty P_i(t)\d t = \frac{\tau^*}{\tau_i}
\end{equation}
In a similar manner, the probability $P_{act}(t)$ that the first
activated depinning event takes place at time $t$ is
\begin{equation}
P_{act}(t) 
=\frac{1}{\tau^*}\exp\left(-\frac{t}{\tau^*}\right)
\end{equation}

Note that the time $t$ and
location $i$ appear as independent variables. When the front is
pinned, the algorithm thus consists of drawing a waiting time from
the distribution $P_{act}(t)$ and to choose the site of depinning
according to the weights $r_i$. The characteristic times involved
in $r_i$ depend on the distribution $P(F_i)$ of individual
depinning forces along the front. In the following we focus on the
two limiting cases of low and high temperature.

At low temperature, when the front is pinned the waiting time is
dominated by the characteristic time of the extremal site. The
sequence of depinning events is identical to  the  one of the
extremal dynamics. The time evolution directly results from the
knowledge of the distribution of depinning forces $P(F_c)$ of the
extremal. At high temperature, the probability of depinning
becomes comparable for all sites along the front.

\subsection{Low Temperature behavior: activated extremal dynamics}

We consider here the case of very low temperature. The loading external
force $F_{ext}$ is constant below the threshold $F^*$. At a given
conformation of the front, all sites satisfying the criterion of
propagation are advanced, the elastic forces are then updated. If
the front is blocked the only site to depin is the extremal one.
The time $\tau$ associated to an iteration is drawn from an
Arrhenius distribution
\begin{equation}
p(\tau)=\frac{1}{\tau}\exp\left(-\frac{\tau}{\tau_c}\right)\;,
\quad \tau_c=
\frac{1}{\nu_0}\max\left[1;\exp\left(\frac{F_{ext}-F_c}{T}\right)\right]
\end{equation}

The knowledge of the distribution of depinning forces $P(F^*-F_c)$
gives us directly the distribution of waiting times in blocked
configurations\cite{SVR-IJMPC02}. The mean waiting time $\langle\tau\rangle$ to
depin from a blocked conformation for an external constant loading
$F_{ext}$ is
\begin{eqnarray}
\label{ext-act}
\langle\tau\rangle &= &
\int_{Fext}^{F^*}\d F_c\exp\left(\frac{F_c-F_{ext}}{T} \right) 
A(F^*-F_c)^{\mu}
\\
&=& T^{\nu(a-1)}\exp \left(\frac{F^*-F_{ext}}{T} \right)
\end{eqnarray}
where we used the condition $T \ll F^*-F_{ext}$ to truncate the
integral in the numerator. In this regime we observe a simple
Arrhenius behavior for the mean depinning time with a power law
correction in the ratio temperature/distance to threshold. On Figure
\ref{act-ext} we show the scaling of this waiting time. After
rescaling by the Arrhenius factor we observe indeed a power law
dependence on the temperature. Numerically we find for the exponent
$\nu(a-1)=2.3$ to compare with the expected value $\nu(a-1)=2.6$ with
$\zeta=1.25$ and $a=2.95$.


\begin{figure}[tbp]
\begin{center}
\epsfig{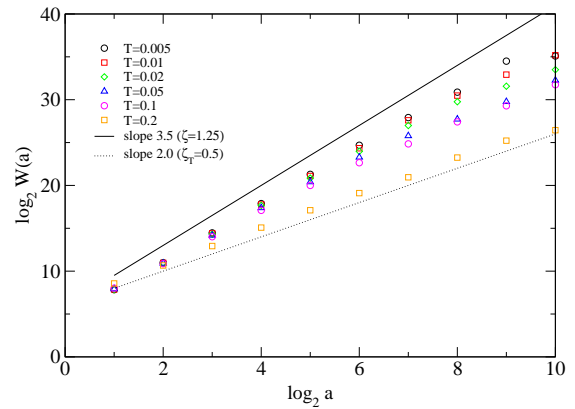}
\caption{\label{wlt_f0} Mean wavelet coefficients obtained for the
depinning front driven at $F_{ext}=0$, ($F^*-F_{ext}=0.569$) for
various temperature. The straight lines indicate the two limit
behaviors of roughness exponents $\zeta=1.25$ and $\zeta_T=0.5$
corresponding to the critical state for temperatures close to zero
a ``SOS-like'' behavior for higher forces. System size $N=4096$,
$32\ 10^6$ iterations.}
\end{center}
\end{figure}


\subsection{High Temperature behavior: Edwards-Wilkinson like dynamics}

At high temperature, all growth probabilities tend to be equal, which
generates an independent evolution at each site.  The result is simply
a white noise morphology, with ever increasing slopes and curvatures.
This regime can only be a transient one.  At long times (and at lower
temperatures), a bias between the growth probability and the curvature
will be felt and hence, this limit will correspond to an annealed
model known as the Edwards-Wilkinson (EW) model\cite{Barabasi-book95},
which indeed give rise to self-affine fronts with a roughness exponent
$\zeta=0.5$.  We thus expect to see a cross-over from $\zeta=1.25$
at small temperature, to $\zeta=0.5$ at larger temperature. This is
precisely what is observed in the numerical simulations as shown in
Figure~\ref{wlt_f0} which represent the scaling behavior of the front
roughness for a same driving force but different temperatures. Here we
measured along the front the averaged wavelet coefficients
$\omega(a)$. For a self-affine front of roughness exponent $\zeta$,
these coefficients are expected to scale as $ \omega(a) \propto
a^{1+2\zeta}$ with the length scale $a$\cite{Simonsen-PRE98}. In
$\log-\log$ scale we observe a scaling behavior with an apparent
exponent decreasing from $\zeta\approx 1.25$ toward $\zeta\approx
0.5$ for growing temperatures.

\begin{figure}[tbp]
\epsfig{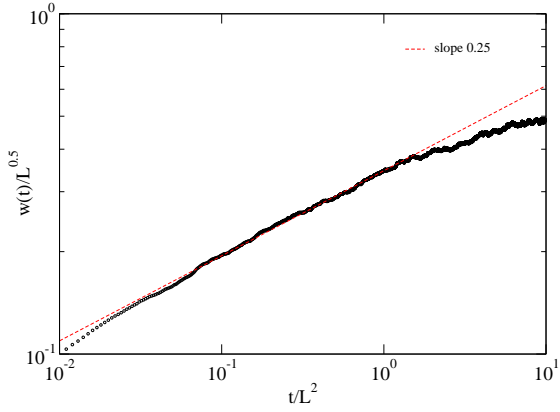}
\caption{\label{EW} Time evolution of the interface width at high
temperature ($T=0.1$) with a low forcing $F^*-F_{ext}=0.39$. The time
and height coordinates are rescaled by $L^z_{EW}$ and $L^{\zeta_{EW}}$
respectively. The value $z_{EW}=2$ and $\zeta_{EW}=0.5$. The evolution
of the width is well described by the power law $w(t)\propto
t^{\beta_{EW}}$ with $\beta_{EW}=0.5$ (dashed lines). These critical
exponents are consistent with a growth of the Edwards-Wilkinson type.}
\end{figure}

To
check that this high-temperature behavior indeed belongs to the EW
universality class, we show on Fig. \ref{EW} the time evolution of the
interface width after rescaling of the time by $L^z$ and $L^\zeta$
with $z=2$ and $\zeta=0.5$. As expected in a EW growth model we
observe a power law behavior with an exponent $\beta=0.25$.

\subsection{Transition from low to high temperature: thermal
 characteristic length}

Let us now consider intermediate temperatures. We can actually
establish a quantitative criterion to discriminate between low and
high temperature behavior.

As stated above the probability of depinning of a
pinned configuration derives from the distribution of depinning
forces along the front. These depinning forces can be ranked from
the smallest (the extremal site) to the largest one. The extremal
site is denoted $i^*$ and the associated depinning force is $F_c$.
The second smallest is $i_2$ and the corresponding force is
$F_{i_2}$.  The typical depinning time $\tau^*$ is:
\begin{equation}
\frac{1}{\tau^*}=\sum_i \frac{1}{\tau_i}=
\frac{1}{\tau_{i^*}}\left(1+ \sum_{j=2}^N
\frac{\tau_{i^*}}{\tau_{i_j}}\right)
\end{equation}
where
\begin{equation}
\frac{\tau_{i^*}}{\tau_{i_j}} = \exp \left( \frac{F_{ext}-F_c}{T}
- \frac{F_{ext}-F_{i_j}}{T} \right) = \exp \left(
\frac{F_{i_j}-F_c}{T} \right)
\end{equation}
We see here that the relative weights corresponding to the
different sites do not depend on the level of external driving but
only on the difference of force to the extremal one. It is thus
clear that as soon as $F_c-F_{i_j} \gg T$ the waiting time
$\tau^*$ is dominated by the extremal time $\tau_{i^*}$. The
ranking allows to restrict to the criterion:
\begin{equation}\label{crit1}
F_c-F_{i_2} \gg T
\end{equation}
The previous
section gave us the knowledge on the distribution of depinning
forces along the front, namely:
\begin{equation}
Q(F^*-F_i)  \propto \frac{1}{L}(F^*-F_i)^{-\varepsilon}
\end{equation}
We may thus estimate the typical difference between the
two smallest depinning forces along the front:
\begin{equation}
\int_{F_c}^{F_{i_2}}\frac{1}{L^D}  (F^*-F)^{-\varepsilon} \d F
=\int_{F^*-F_{i_2}}^{F^*-F_c}  \frac{1}{L^D}x^{-\varepsilon} \d x \approx
\frac{1}{L^D}
\end{equation}
 where $1/L^D$ in the integral is a normalizing
factor of the subcritical force distribution and $L^D$ at the right hand sides is the number of subcritical site. A Taylor expansion around
$F_c$ leads to
\begin{equation}
(F_c-F_{i_2}) \approx (F^*-F_c)^\varepsilon
\end{equation}

\begin{figure}[tbp]
\begin{center}
\epsfig{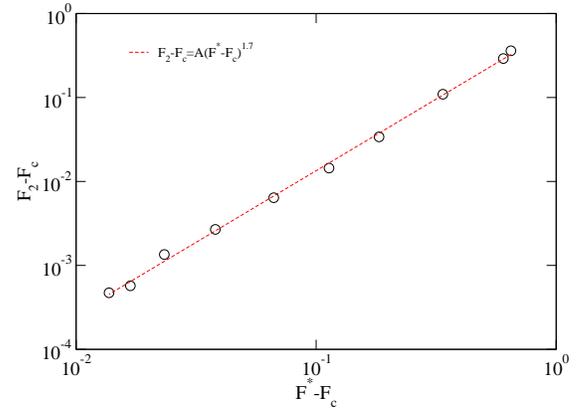}
\end{center}
\caption{\label{f2vsfc} Difference between the two smallest depinning
forces $F_2-F_c$ along the front against the difference $F^*-F_c$. The
data are well described by power law of exponent $\varepsilon=0.7$.}
\end{figure}

We see in Fig. \ref{f2vsfc} that this scaling is nicely observed with
the value $\varepsilon=1.7$ consistent with the estimation obtained
above for the distribution of subcritical depinning forces.
The extremal value of the depinning force $F_c$ is a fluctuating
quantity during the course of the propagation. In the first part
of this paper we saw that the difference $F^*-F_c$ can be
associated to a typical distance $d$ between consecutive depinning
sites by the scaling $d \propto (F^*-F_c)^{-\nu}$.  The criterion
for extremal depinning during the propagation can thus be
characterized by a distance $\xi_T$ below which the activated site
is the extremal one and above which the activated site is randomly
chosen along the front:
\begin{equation}
 \xi_T \propto T^{-\frac{\nu}{ \varepsilon}}
\end{equation}

\begin{figure}[tbp]
\begin{center}
\epsfig{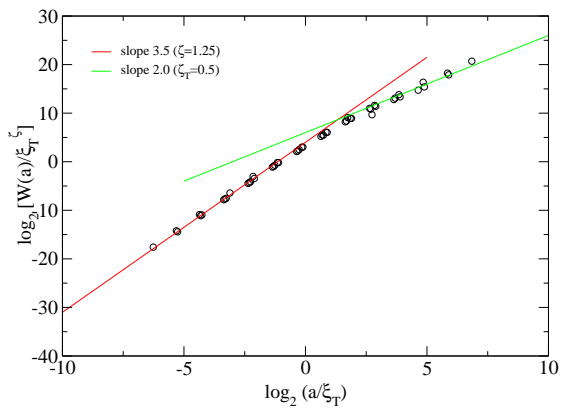}
 \caption{\label{wlt_f0_collapse}After rescaling the
abscissa by the characteristic length $\xi_T\propto
T^{-\nu/\epsilon}$ with $\nu=1.33$ and $\epsilon=1.4$, and the
ordinate by $\xi_T^\zeta$ with $\zeta=1.25$, the mean wavelet
coefficients obtained for the depinning front driven at the
constant driving force $F_{ext}=0$, ($F^*-F_{ext}=0.569$) for
different temperatures collapse onto a single master curve.}
\end{center}
\end{figure}

In Fig. \ref{wlt_f0}, we presented the scaling of the front roughness
obtained for a constant forcing at various temperatures. As expected,
the scaling evolves from the zero temperature behavior $\zeta=1.25$ to an
annealed behavior $\zeta=0.5$: the various curves are distributed
between  these two limit cases. However, rescaling the lateral
length by the temperature dependent characteristic length $\xi_T$, we
can collapse all results onto a single master curve (see
Fig. \ref{wlt_f0_collapse}) where the small arguments correspond to
the critical behavior and the large arguments to the annealed
behavior.

\section{Conclusion}

We studied the depinning of an elastic line in strong pinning
conditions. 

In a first part we focused on the critical behavior at zero
temperature, based on an extremal dynamics to perform numerical
simulations. Beyond the scaling properties of the front roughness and
the avalanche dynamics observed at the depinning transition our main
results concern the statistics of the depinning forces. At each
conformation of the front we can associate the minimum force $F_c(t)$
needed to depin the front. The distribution of these depinning forces
present a singular behavior in the vicinity of its maximum value, the
critical threshold $F^*$ is $P(F_c)\propto (F^*-F_c)^\mu$ where the
critical exponent $\mu$ directly depends on the roughness exponent
$\zeta$ of the depinning front. Moreover this distribution relative to
large jumps $d$ in the activity adopts a universal form for all values
of $F$.

 Moreover it appeared that the critical dynamics concentrates on a
very tenuous set of sites (termed subcritical) along the front, in
fact a fractal structure. Beyond neighboring effects, the critical
site is always chosen among this set. The latter can thus be regarded
as precursors of the future avalanches. All of these sites are
characterized by a sub-critical depinning force, namely the force $F$
needed to make them move is below the critical threshold $F^*$. Again
the distribution of the depinning forces associated to these
precursory sites presents a universal character: $P(F)\propto
L^{-1}(F^*-F)^{-\epsilon}$, again composed of a universal form when
conditioned by the distance to the active site.

In a second part we extended the study of the depinning transition at
finite temperature. The front was driven at a constant below threshold
driving force. A simple Monte-Carlo algorithm was developed to
simulate the activated dynamics. 

Two limit cases were considered. At very low temperature or at small
scale, the extremal site is by far the most probable site to depin and
all critical properties are preserved. In this limit, the finite
temperature then mimics the extremal dynamics. Assuming an Arrhenius
behavior for the depinning from any individual trap we are able to
associate a physical time to this quasi extremal dynamics. The
knowledge of the depinning force distribution $P(F_c)\propto
(F^*-F_c)^\mu$ then allows a quantitative estimate of the front
velocity. We recover a simple Arrhenius dependence with a temperature
power law dependence.
At high temperature or at large scales, all sites can
equally be activated. This annealed dynamics has been shown to be well
described by an Edwards-Wilkinson growth model.  The transition
between these two regimes can be estimated using the ratio of the
temperature to the gap between the two smaller depinning forces. We
showed that this gap can be naturally associated to a temperature
dependent characteristic length along the front $\xi_T \propto
T^{-1/\varepsilon \nu}$. Below the characteristic length $\xi$ the
front adopts naturally a critical dynamics, at larger scales the
annealed dynamics dominates.

\begin{appendix}
\section{A solvable mean field depinning model}

We present in this appendix a mean field variant of the extremal
depinning model studied in section II. We first note that we can
set all thresholds to 0, and provided that the moves are still random,
the behavior remains similar. In this context the extremal site $i^*$ simply
minimizes the local elastic force:

\be
F_c= F^{el}_{i^*}=\min_i F^{el}_i \;.
\ee

Once the extremal site has been determined the front advances up to
the next trap. However we introduce here a change in the definition of
the elastic forces. In the original Laplacian version
$F^{el}_i=-h_{i-1} +h_i -h_{i+1}$, the extremal site is unloaded
proportionally to the local displacement just covered and conversely
its two nearest neighbors are equally loaded. In the mean field
version we present here, the two sites loaded to keep the elastic
force balance are chosen randomly along the front, hence the term mean field.

Let us call $x$ the gap between the local elastic force and the
critical threshold, $x=F^{el}-f^*$, $p(x)$ its distribution density,
and $P(x)$ the cumulative probability $P(x)=\int_x^\infty p(y) dy$
(from which $p(x)=-P'(x)$ results).

For a system of size $L$, the smallest $x$ has a density $p_s(x)$
and cumulative $P_s(x)$.  We have $P_s(x)=P(x)^L$, and thus
$p_s(x)=LP(x)^{L-1} p(x)$.

Ignoring spatial correlations, we can write a mean-field master
equation for $p(x)$. 

 \be\ba{lll} \displaystyle \frac{dp(x)}{dt}
&\displaystyle=&-p_s(x)-2p(x)\\
&&\displaystyle +(1/2)\int_0^2 p_s(x-y) dy
\\&&\displaystyle +2\int_0^1 p(x+y) dy
\\ 
&\displaystyle=&-p_s(x)-2p(x)\\
&&\displaystyle +(1/2)(P_s(x-2)-P_s(x))\\
&&\displaystyle +2(P(x)-P(x+1))
\ea\ee 

 The terms of the
right hand side correspond to the probability of occurrence of the
following events {\it i)} $x$ is the extremal value of the elastic
force along the site; {\it ii)} Before the update, one of the two
random sites to be loaded carried a force $x$; {\it iii)} After update
the advancing site carries a force $x$; {\it iv)} After update one of
the two random sites carries a force $x$.

  In the steady state, we have
  \be\label{eq:full}
  -2P'(x)-P'_s(x)=(1/2)(P_s(x-2)-P_s(x))+2(P(x)-P(x+1))
  \ee

In order to integrate this equation, we distinguish in the following
between three different intervals. The region $1$ $(x<0)$ corresponds
to the subcritical sites. The region $3$ $(x>1)$ corresponds to deeply
pinned sites which have no chance to become neither subcritical nor
critical at the next depinning event. The intermediate region $2$
$(x>1)$ exchanges with the two other ones.  For $x<0$, we expect
$P(x)$ to be equal to $1$ but a small correction vanishing as $L$
diverges.  $P(x)=1-\Psi_L(x)$. We look for $\Psi_L(x)=(1/L)\Phi(x)$.
Thus $P_s(x)=(1-(1/L)\Phi(x))^L\to \exp(-\Phi(x))$.  The equations
corresponding to the three regions write:

\begin{itemize}
\item Part 1:
\be
-P'_s(x)=(1/2)(1-P_s(x))+2(1-P(x+1))\;,\quad (x<0)
\ee

\item  Part 2:
  \be
  -P'(x)=P(x)-P(x+1)+(1/4)\;,\quad (0<x<1)
  \ee

\item  Part 3:
  \be
  -2P'(x)=(1/2)P_s(x-2)+2P(x)\;,\quad (x>1)
  \ee

\end{itemize}
 and the solutions are respectively
\begin{itemize}

\item  Part 1: $(x<0)$
\begin{eqnarray}
P_s(x)&=&-x^2-2x\;,\\
P(x)&=&1-\frac{1}{L}\frac{2(x+1)}{x(x+2)}\;,
\end{eqnarray}

\item  Part 2 and 3: $(x>0)$
\begin{equation}
P(x)=(1/4)(x-2)^2 \;.
\end{equation}

\end{itemize}

These analytical expressions are compared with numerical results on
Fig. \ref{fi:mf}. Beyond the excellent agreement between analytics and
numerics, the similarity between Fig. \ref{histo_f} and \ref{fi:mf} shows that
this simple mean field model allow to recover the main feature of the
original model. The distribution of local depinning forces is composed
of a size independent supercritical part and of a size dependent
subcritical part which presents in addition a singular behavior close
to the forces approach the critical threshold.

\end{appendix}

\bibliography{vdb,depinning}

\end{document}